**Schoolyard Greening, Child Health, and Neighborhood Change: A Comparative Study of Urban U.S. Cities**

Mahshid Gorjian

Department, University of Colorado Denver, Denver, USA.

Contributing authors: mahshid.gorjian@ucdenver.edu

OCID: 0009-0000-9135-0687

**Abstract**

**Background:** Schoolyard greening has emerged as an innovative approach to enhancing children's health and environmental equity in urban U.S. cities. Yet, the implications for neighborhood dynamics and social equity are insufficiently understood.

**Methods:** This comparative literature review synthesizes quantitative and qualitative evidence from peer-reviewed studies and case analyses of schoolyard greening in major U.S. cities.

**Results:** Schoolyard greening consistently increases utilization and has a positive, though variable, effect on children's physical activity and well-being. However, the benefits are not always equitably distributed, and greening projects can catalyze neighborhood change, sometimes leading to green gentrification.

**Conclusions:** Greening urban schoolyards offers benefits for children's health and urban sustainability but poses challenges for equity and social justice. Policies must prioritize inclusive, community-driven approaches to ensure the gains of greening are shared by vulnerable populations.

**Background**

Childhood obesity and physical inactivity are critical public health issues in the United States, disproportionately affecting children in low-income and minority communities (Anthamatten et al., 2011). As environmental factors have become recognized as central to these disparities, policy and design interventions targeting the built environment particularly schoolyards have garnered increasing attention (Anthamatten et al., 2014). Well-designed schoolyards are posited not only to increase children's opportunities for physical activity but also to support broader goals of social equity, community health, and environmental justice (Bikomeye et al., 2021; Wolch et al., 2014).

**Schoolyard Greening in Urban Contexts**

Urban schoolyards in the United States are frequently characterized by limited vegetation, aging infrastructure, and few amenities conducive to active or imaginative play. This is especially pronounced in schools serving disadvantaged populations (Anthamatten et al., 2011). In response, a variety of greening interventions have been implemented, ranging from the installation of gardens and trees to comprehensive renovations featuring art, play structures, and natural elements (Anthamatten et al., 2014; Raney et al., 2023).

**Theoretical Framework and Rationale**

The link between the built environment and physical activity is well-established. Access to green spaces, play equipment, and safe recreational facilities has been associated with higher levels of activity among children (Anthamatten et al., 2011; Bikomeye et al., 2021). Schoolyard greening aims to address deficits in these resources while simultaneously advancing community development objectives (Wolch et al., 2014).

**Research Questions:**

1. What are the impacts of schoolyard greening on children's physical activity and well-being?

2. How do schoolyard greening projects affect patterns of neighborhood change, including risks of green gentrification?

3. Are the benefits of schoolyard greening equitably distributed among diverse urban populations?

**Methods**

This study adopts a comparative literature review methodology, synthesizing findings from quantitative and qualitative studies of schoolyard greening in major U.S. cities. The review includes peer-reviewed journal articles, systematic reviews, and case studies published between 2010 and 2023, focusing on relationships between greening interventions, child health outcomes, and neighborhood change (Anthamatten et al., 2011; Anthamatten et al., 2014; Bikomeye et al., 2021; Raney et al., 2023). Studies using validated measures such as the System for Observing Play and Leisure Activity in Youth

**Results**

**Utilization and Physical Activity**

Renovated and greened schoolyards consistently show increased levels of utilization compared to un-renovated controls. For example, the Learning Landscapes program in Denver, Colorado, which introduced culturally relevant and nature-based elements into schoolyards, led to significantly higher use, particularly among girls during before- and after-school times (Anthamatten et al., 2011). However, there were no statistically significant differences in the rate of physical activity i.e., the proportion of active children between renovated and un-renovated schoolyards, suggesting that while more children are attracted to greened schoolyards, the likelihood of engaging in moderate or vigorous activity may depend on factors such as programmatic support and supervision.

Systematic reviews confirm these findings, highlighting that greening interventions promote overall physical activity, support psychological well-being, and enhance cognitive outcomes (Bikomeye et al., 2021). Play zone diversity and natural features, such as gardens and trees, also encourage unstructured, inclusive play among diverse groups of children (Raney et al., 2023).

**Equity and Environmental Justice**

Despite their promise, the benefits of schoolyard greening are not always equitably realized. Children in low-income and minority communities often continue to face barriers to high-quality recreational environments, and parental perceptions of safety and accessibility remain critical determinants of use (Anthamatten et al., 2011). Interventions that integrate culturally relevant features and involve community stakeholders have been shown to improve utilization and reduce disparities (Anthamatten et al., 2014).

**Neighborhood Change and Green Gentrification**

A growing body of evidence links urban greening projects, including schoolyard renovations, to broader neighborhood change processes, sometimes resulting in "green gentrification" (Anguelovski et al., 2019; Gould & Lewis, 2017). While environmental improvements can attract investment and enhance urban sustainability, they can also drive-up property values and displace long-term, often vulnerable residents, potentially undermining the very goals of health equity and environmental justice that motivated the greening interventions (Wolch et al., 2014). Efforts to make cities "just green enough" aim to balance environmental enhancements with protections against displacement and exclusion.

**Discussion**

This synthesis demonstrates that schoolyard greening is a promising strategy to support child health and address disparities in urban environments. Increased utilization of schoolyards, particularly by girls and children in low-income neighborhoods, is a consistent outcome of greening initiatives (Anthamatten et al., 2011; Anthamatten et al., 2014). However, the variability in physical activity rates points to the importance of complementary interventions, such as programmatic activities and community engagement.

At the same time, schoolyard greening must be approached with caution regarding its potential to trigger green gentrification and contribute to the displacement of vulnerable populations (Anguelovski et al., 2019). Achieving environmental justice requires that greening efforts are paired with policies to preserve affordable housing and ensure community participation throughout the planning and implementation processes (Gould & Lewis, 2017; Wolch et al., 2014).

**Study Limitations**

This review is limited by reliance on existing studies, which may be influenced by context-specific factors. In addition, there is a lack of longitudinal data on the long-term impacts of schoolyard greening. Most available research has focused on utilization and immediate health outcomes, with less attention to the long-term social and demographic effects of neighborhood change.

**Conclusions**

Schoolyard greening initiatives in urban U.S. cities offer significant promise for enhancing children's physical activity, well-being, and environmental equity. However, without intentional, justice-centered approaches, these projects may inadvertently contribute to the exclusion of the populations they are intended to benefit. Future research should prioritize longitudinal, community-engaged studies to elucidate the mechanisms of sustained impact and to develop policies that ensure greening benefits are shared by all.

**References**

Anthamatten, P., Brink, L., Lampe, S., Greenwood, E., Kingston, B., & Nigg, C. (2014). An assessment of schoolyard renovation strategies to encourage children's physical activity. *International Journal of Behavioral Nutrition and Physical Activity, 11*, 27. https://doi.org/10.1186/1479-5868-11-27

Anthamatten, P., Fiene, E., Kutchman, E., Mainar, M., Brink, L., Lampe, S., & Nigg, C. (2011). Playground spaces and physical activity: Investigating perceptions of schoolyard environments among children in Denver. *International Journal of Behavioral Nutrition and Physical Activity, 8*, 34. https://doi.org/10.1186/1479-5868-8-34

Bikomeye, J. C., Balza, J., & Beyer, K. M. (2021). The Impact of Schoolyard Greening on Children's Physical Activity and Socioemotional Health: A Systematic Review of Experimental Studies. *International journal of environmental research and public health*, *18*(2), 535. https://doi.org/10.3390/ijerph18020535

Gorjian, (2025). Greening Schoolyards and the Spatial Distribution of Property Values in Denver, Colorado (Version 1). figshare. https://doi.org/10.6084/m9.figshare.29539853.v1

Raney, M. A., Daniel, E., & Jack, N. (2023). Impact of urban schoolyard play zone diversity and nature-based design features on unstructured recess play behaviors. *Landscape and Urban Planning, 230*, 104632. https://doi.org/10.1016/j.landurbplan.2022.104632

Anguelovski, I., Connolly, J. J. T., Pearsall, H., Shokry, G., Checker, M., Maantay, J., Gould, K., Lewis, T., Maroko, A., & Roberts, J. T. (2019). Why green "climate gentrification" threatens poor and vulnerable populations. Proceedings of the National Academy of Sciences of the United States of America, 116(52), 26139-26143.

Wolch, J. R., Byrne, J., & Newell, J. P. (2014). Urban green space, public health, and environmental justice: The challenge of making cities 'just green enough'. *Landscape and Urban Planning, 125*, 234–244. https://doi.org/10.1016/j.landurbplan.2014.01.017

Raina, A. S., Mone, V., Gorjian, M., Quek, F., Sueda, S., & Krishna- murthy, V. R. (2024). Blended physical-digital kinesthetic feedback for mixed reality-based conceptual design-in-context. In GI '24: Proceedings of the 50th Graphics Interface Conference (Article 6, pp. 1–16). ACM. https://doi.org/10.1145/3670947.3670967

Li, J., Ossokina, I. V., & Arentze, T. A. (2024). The Impact of Urban Green Space on Housing Value: A Combined Hedonic Price Analysis and Land Use Modeling Approach. *Journal of Sustainable Real Estate, 16*(1). https://doi.org/10.1080/19498276.2024.2432758